\documentstyle[prl,aps,floats,epsf]{revtex}
\begin{document}
\twocolumn[
\hsize\textwidth\columnwidth\hsize\csname@twocolumnfalse\endcsname

\draft
\title{The Neutron Peak in the Interlayer Tunneling Model of High Temperature
Superconductors}
\author{Lan Yin$^a$, Sudip Chakravarty$^a$, and Philip W. Anderson$^b$}
\address{$^a$Department of Physics and Astronomy, University of California, Los Angeles, CA 90095\\
$^b$Joseph Henry Laboratories of Physics, Princeton
University, Princeton, NJ 08544}
\date{\today}
\maketitle
\begin{abstract}
Recent neutron scattering experiments in YBCO exhibit an unusual magnetic peak that appears only below the superconducting transition temperature. The experimental observations are explained within the context of the interlayer tunneling theory of high temperature superconductors.
\end{abstract}
]
Many experimental observations in high temperature superconductors are commonly
fit with a phenomenological model that derives from the original theory of
Bardeen, Cooper and Schrieffer (BCS). The model has certain ingredients. It is
characterized by a gap equation corresponding to a presumed symmetry of the
order parameter,  with an adjustable dimensionless coupling constant, or
alternately an adjustable ratio of the gap to the superconducting transition
temperature, and a given Fermi surface. Inherent in this description are the
coherence factors that determine a number of interference processes unique to
the BCS theory. Although a microscopic derivation of this effective model for
the high temperature superconductors does not exist, the model is still used with
a considerable degree of confidence. The main difficulties, to which we return
below, are the unusual normal state properties of these materials and the
enormously high transition temperatures, but there are many others. Such
difficulties are extensively surveyed in the literature\cite{Anderson1}. 

In the absence of a microscopic derivation, it is
useful to ask if this phenomenological BCS model is unique and if an alternative
phenomenological model exists that is capable of capturing features of
these superconductors. One such physically motivated model was elaborated in a recent
paper\cite{Chakravarty1}.
In the present paper we examine this model, called the interlayer tunneling
model, to interpret the startling neutron scattering experiments in optimally
doped YBCO\cite{Keimer1}. 

These neutron scattering experiments are startling for a number of reasons.
The experiments appear to have established that there are no sharp, or even
broad, features in the magnetic excitation spectrum in the normal state. In
contrast, the superconducting state exhibits a sharp magnetic feature that
appears to be localized both in energy and momentum. It is located at an energy
of 41 meV and near a wavevector $(\pi/a, \pi/a, \pi/c_b)$, where $a$ is the
lattice spacing of the square-planar CuO lattice, and $c_b$ is the distance
between the nearest-neighbor copper atoms within a bilayer. The intensity under
the peak vanishes at the transition temperature, but its frequency softens very
little.

In principle, there can be many explanations, and such explanations have been
proposed\cite{Other}. None of these explanations are fully microscopic, nor are
they fully consistent with some of the other prominent experimental observations
in these materials. In particular, we draw attention to the fact that
explanations that rely on forming {\em coherent} linear superpositions of the
bilayer bands would also predict large splittings of the bands in certain
regions of the Brillouin zone. Such splittings, studied extensively in 
experiments, are missing to a high degree of accuracy\cite{Photo}. The argument
that the observed bands are highly renormalized has little force. If the
renormalization effects are so strong as to reduce the splitting to nearly zero,
it is not meaningful to speak of a coherent superposition of states at the
experimentally relevant temperatures.

The explanation we have to offer within the interlayer tunneling model is
exceedingly simple and purely kinematic in origin.  We also 
present illustrative but quantitative results to compare with experiments. The comparison shows
that many of the features of the experiments are captured well.
While many improvements can possibly be made, the
kinematic aspects should survive future scrutiny. 

The model Hamiltonian, motivated earlier\cite{Chakravarty1}, is that of a
bilayer complex. Its generalization to many layers is straightforward and not
essential for the present purpose. It is
\begin{eqnarray}
H&=&\sum_{{\bf k}\sigma i}\varepsilon_{\bf k}c_{{\bf k}\sigma i}^{\dagger}c_{{\bf
k}\sigma i}\nonumber \\
&-&
\sum_{{\bf q},{\bf k},{\bf k}', \sigma,\sigma',i}V_{{\bf q},{\bf k},{\bf
k}'}c_{{\bf k}\sigma i}^{\dagger}c_{-{\bf k}+{\bf q}\sigma' i}^{\dagger}c_{-{\bf
k}'+{\bf q}\sigma' i}c_{{\bf k}'\sigma i}\nonumber \\
 &-&\sum_{{\bf q},{\bf k},\sigma,\sigma',i\ne
j}T_J({\bf q},{\bf k})\left[c_{{\bf k}\sigma i}^{\dagger}c_{-{\bf
k}+{\bf q}\sigma' i}^{\dagger}c_{-{\bf k}+{\bf q}\sigma' j}c_{{\bf k}\sigma
j}+{\rm h. c.}\right].
\end{eqnarray}
Here $ i=1,2$ is the layer index. The fermion operators are labeled by the
spin $\sigma$ and the wavevector ${\bf k}$; $V$ is  the in-plane pairing interaction. The
last term describes tunneling of pairs between the layers. Such a Hamiltonian should be
understood from the point of view of an effective Hamiltonian, as emphasized
recently\cite{Lan}. It incorporates the unique feature of the interlayer
mechanism, that tunneling occurs with conservation of transverse momentun $\bf
k$. Therefore, the momentum sum in the $T_J$ term is only over $\bf k$ and $\bf
q$. Disorder between the layers is weak, and even disorder would not affect
this crucial difference between the tunneling and interaction terms. 

Only in the
subspace in which both the states
$({\bf k}\uparrow)$ and $(-{\bf k}\downarrow)$ are both simultaneously occupied
or unoccupied, is the following reduced Hamiltonian  sufficient:
\begin{eqnarray}
H_{\rm red}&=&\sum_{{\bf k}\sigma i}\varepsilon_{\bf k}c_{{\bf k}\sigma
i}^{\dagger}c_{{\bf k}\sigma i}-
\sum_{{\bf k},{\bf k}',i}V_{{\bf k},{\bf k}'}c_{{\bf k}\uparrow
i}^{\dagger}c_{-{\bf k}\downarrow i}^{\dagger}c_{-{\bf k}'\downarrow i}c_{{\bf k}'\uparrow
i}\nonumber \\
& -&\sum_{{\bf k},i\ne
j}T_J({\bf k})\left[c_{{\bf k}\uparrow i}^{\dagger}c_{-{\bf k}\downarrow
i}^{\dagger}c_{-{\bf k}\downarrow j}c_{{\bf k}\uparrow j}+{\rm h. c.}\right].
\label{HRED}
\end{eqnarray}
This is because these are the only matrix elements that survive. It must be
remembered that the actual Hamiltonian is $H$ and that $H_{\rm red}$ is merely a
convenient tool to calculate the matrix elements in the paired subspace. We do
not anticipate that the Schrieffer zero-momentum pairing hypothesis will be
necessarily as accurate in this case as in BCS, but it surely will be a guide.

An important feature is the
absence of the single particle tunneling term from the Hamiltonian, which
allows for a coherent superposition of the states of
the layers that leads to a splitting, not observed in experiments. Therefore,
such terms are not included in our model.
Although the single particle tunneling term is absent, incoherence does
not preclude a particle-hole two particle tunneling
term\cite{Chakravarty1,Lan}, because the two particle term is generated by a
second order virtual process and the question about coherence is irrelevant; the
energy need not be conserved in a virtual process. However, this term does not
appear  to be
crucially important for optimally doped materials (except, see later). For
underdoped materials exhibiting ``spin gap" phenomena\cite{Spingap}, this
neglected term can be quite important\cite{Anderson2}. We also note that we
shall treat the Fermion operators as ordinary anticommuting fermion operators.
This can only be a crude approximation due to the non-Fermi liquid feature of
the normal state. A possible improvement along the lines
discussed earlier\cite{Lan} is exceedingly complex. 

The mean field treatment of the reduced Hamiltonian is
straightforward\cite{Chakravarty1} and leads to the gap equation:
\begin{equation}
\Delta_{\bf k}={1\over 1-\chi_{\bf k}T_J({\bf k})}\sum_{{\bf
k}'}V_{{\bf k},{\bf k}'}\chi_{{\bf k}'}\Delta_{{\bf k}'},
\end{equation}
where $\chi_{\bf k}=(\Delta_{\bf k}/2E_{\bf k})\tanh(E_{\bf k}/2T)$ is the pair
susceptibility, with 
$E_{\bf k}=\sqrt{(\varepsilon_{\bf k}-\mu)^2+\Delta_{\bf k}^2}$. Here $\mu$ is the
chemical potential. Note that, until now, we have not specified the symmetry of 
the in-plane
pairing kernel. We shall now assume it to be of the symmetry $d_{x^2-y^2}$,
$V_{{\bf k},{\bf k}'}=Vg_{\bf k}g_{{\bf k}'}$, where
$g_{\bf k}={1\over 2}\left[\cos (k_xa)-\cos (k_ya)\right]$.

Further specifications are necessary. The in-plane one electron dispersion is
chosen to be
$\varepsilon_{\bf k}=-2t\left[\cos (k_xa)+\cos (k_ya)\right]
+4t'\cos (k_xa)\cos (k_ya)$.
The quantitative calculations are carried out with a representative set of
parameters appropriate to YBCO. These are $t=0.25$ eV, $t'=0.45t$, and
$\mu=-0.315$ eV, corresponding to an open Fermi surface, with a
band filling of 0.86. The quantity
$T_J({\bf k})$ was argued\cite{Chakravarty1} to be 
$T_J({\bf k})=(T_J/16)\left[\cos (k_xa)-\cos (k_ya)\right]^4$.
The validity of this expression is now  strengthened by detailed
electronic structure calculations\cite{Andersen} for the single particle
hopping matrix element $t_{\perp}({\bf k})$. 

The magnetic neutron scattering intensity is proportional to the imaginary part of the 
wavevector and the frequency dependent spin susceptibility $\chi({\bf q},\omega)$, which
for the above model is simply the expression:
\begin{eqnarray}
\chi({\bf q},\omega)&=&\sum_{\bf k}\Bigg[{A^+_{{\bf k},{\bf q}}F^-_{{\bf k},{\bf
q}}\over \Omega^1_{{\bf k},{\bf q}}(\omega)}\nonumber \\
&+&{A^-_{{\bf k},{\bf q}}
(1-F^+_{{\bf k},{\bf q}})\over 2}\left({1\over \Omega^{2+}_{{\bf k},{\bf
q}}(\omega)}- {1\over \Omega^{2-}_{{\bf k},{\bf q}}(\omega)}\right)\Bigg],
\end{eqnarray}

where 
\begin{equation}
A^{\pm}_{{\bf k},{\bf q}}={1\over 2}\left[1\pm{(\varepsilon_{\bf
k}-\mu)(\varepsilon_{{\bf k}+{\bf q}}-\mu)+
\Delta_{\bf k}\Delta_{{\bf k}+{\bf q}}\over E_{\bf k}E_{{\bf k}+{\bf
q}}}\right],
\end{equation}
$\Omega^1_{{\bf k},{\bf q}}(\omega)=\omega - (E_{{\bf k}+{\bf q}} -E_{\bf
k})+i\delta$,
$\Omega^{2\pm}_{{\bf k},{\bf q}}(\omega)=\omega\pm (E_{{\bf k}+{\bf q}} +E_{\bf
k})+i\delta$, $F^{\pm}_{{\bf k},{\bf q}}=f(E_{{\bf
k}+{\bf q}})\pm f(E_{\bf k})$, and $f(x)$ is the Fermi function.

Note that near $T=0$ only the $A^-$ terms contribute, and these are negiligible
unless $\Delta_{\bf k}$ and  $\Delta_{\bf k+q}$ are of opposite sign, as noted
in Ref.~\cite{Keimer1}. In fact, nothing is observed for $q\approx 0$ or $2\pi$,
as expected, and the peak appears at ${\bf q}=(\pi/a,\pi/a)$, connecting two
opposite-sign lobes of $x^2-y^2$. At higher temperatures, the $A^+$ part is 
temperature dependent, but almost frequency independent function for
experimentally relevant frequencies. The $\chi$ with the $A^+$ term omitted
will be denoted by $\bar \chi$. We shall first calculate the imaginary part of
this susceptibility by  restricting all wavevectors to be in the CuO-plane.
Later, we shall discuss the dependence on the momentum transfer perpendicular to
the plane.

\begin{figure}[htb]
\centerline{\epsfxsize= 2.4 in \epsffile{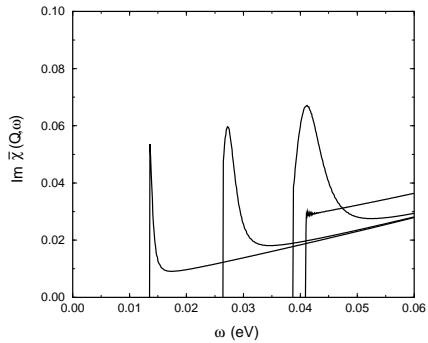}}
\caption{Im $\bar\chi$ at $T=0$. 
From left to right, $T_J=$0.025, 0.05, 0.075 eV. The curve
corresponding to a step discontinuity at the edge is for BCS, with the maximum
$d$-wave gap equal to 0.025 eV.}
\label{ZeroT}
\end{figure}

At $T=0$, Im $\bar \chi$  is calculated by solving the gap equation for a set of $T_J$, and with $\lambda\equiv N(0)V=0.184$, where $N(0)$ is the density of states per spin at the Fermi energy ($V=0.2$ eV, $N(0)=0.92$). These
are shown in Fig.~\ref{ZeroT}. 
While the calculations show a peak at the threshold for $T_J\ne
0$, in the pure BCS case, $T_J=0$, the intensity is a step discontinuity at the
threshold, and other excitonic enhancement mechanisms are necessary to produce a
peak at the threshold. 

\begin{figure}[htb]
\centerline{\epsfxsize=2.4 in \epsffile{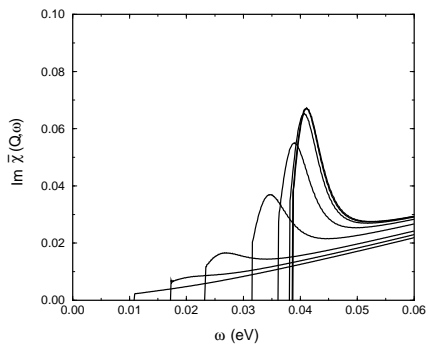}}
\caption{Im $\bar\chi$ for $T_J=0.075$ eV. 
From left to right, $T=$120, 110, 100, 80, 60, 40, 20, and 0 K. The
results for $T=0$ and 20 K are almost indistinguishable.}
\label{FiniteT}
\end{figure}

\begin{figure}[htb]
\centerline{\epsfxsize=2.4 in \epsffile{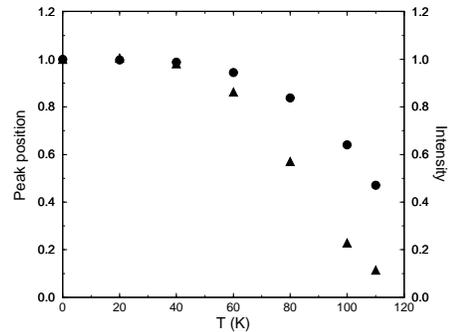}}
\caption{The intensity (solid triangles) and the position of the peak (filled circles) normalized to the zero temperature values. The position of the peak is at 41.2 meV at $T=0$.}
\label{Int}
\end{figure}

For illustrative purposes the  gap equation is solved  at finite temperatures for $\lambda=0.184$ and for a 
fixed $T_J=0.075$ eV. No attempts were made to fit precisely the experimental data. The results are plotted in Fig. ~\ref{FiniteT}.
We calculate the intensity under  the
peak and plot it as a function of temperature. The intensity was calculated by fitting a linear background at high energies.
The results are shown in Fig.~\ref{Int}.
The intensity falls off to zero as the
temperature is raised to $T_c$ and the results compare well with experiments.
Also shown in Fig.~\ref{Int}, is the position of the peak, whose softening is 
weaker than the fall-off of the intensity. However, the experimental softening
of the position of the peak is  even weaker.

The question to address is why Im $\bar \chi$
produces a peak in the interlayer tunneling model, while the BCS model exhibits
only a step discontinuity at the threshold. The reason is simple kinematics
that holds for an open Fermi surface combined with an unusual feature of the
interlayer gap equation. This feature was emphasized in early works on the gap
equation\cite{AndersonA}, but has not been specifically noted in recent
calculations: it is that when $T_J$ dominates, as is necessarily so for high $T_c$, the
density of states for a given value of $T_{J}/2$ is approximately a
$\delta$-function at $T_{J}/2$. Concomitant with this is the fact that
electrons with $|\varepsilon_{\bf k}-\mu|>T_{J}({\bf k})/2$ are unaffected by pairing, in
contrast to BCS where effects of the gap extend out to high energies.

For simplicity, consider $T=0$, set the coherence factor to
unity and look at the wavevector ${\bf Q}=(\pi/a,
\pi/a)$. In the BCS case, that is with $T_J({\bf k})$ set equal to zero,
$(E_{\bf k}+E_{{\bf Q}+{\bf k}})$ has either minima, or saddle points. The minima
give rise to a step discontinuity at the threshold at approximately twice  the maximum of the
$d$-wave gap (See Fig.~\ref{ZeroT}.). The saddle points give rise to van Hove singularities which are
generally not at the threshold, but can be brought close to the threshold by
adjusting parameters.  

\begin{figure}[htb]
\centerline{\epsfxsize=2.4 in \epsffile{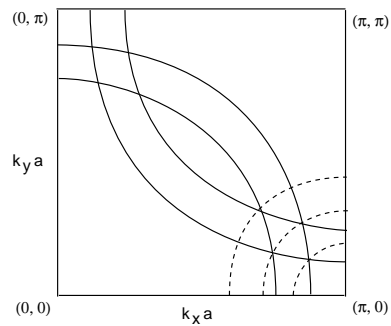}}
\caption{The superconducting region and its mapping under ${\bf k}\to {\bf
Q-k}$. The diamond shaped overlap regions contribute to the peak of the
imaginary part of the spin susceptibility. The dashed arcs are the constant
$\omega$-contours centered at $(\pi,0)$.}
\label{FS}
\end{figure}

Now consider the case of very small in-plane pairing kernel in the
interlayer tunneling equation. The unusual feature is that the superconducting region is a very narrow region around the Fermi line as shown in
Fig.~\ref{FS}. 
The scattering surface is now radically different.  The
contribution to the peak must come from regions in which both
$\Delta_{\bf k}$ and $\Delta_{\bf k+Q}$ are finite. Also, note that
$E_{\bf k+Q}=E_{\bf Q-k}$. The image of the superconducting region including the
Fermi line under the mapping ${\bf k}\to {\bf Q-k}$ is shown in Fig.~\ref{FS}.
The regions in which both $\Delta_{\bf k}$ and $\Delta_{\bf Q-k}$ are finite are
the diamond shaped overlap regions, with one of its diagonals given by
$k_x+k_y=\pi$. Consider the lower diamond. In this region, we can write
$E_{\bf k}\approx T_J\cos^4(d/\sqrt{2})$, where $d$ is the distance from $(\pi/a, 0)$. The corresponding constant $\omega$-contours are the arcs shown in Fig.~\ref{FS}.

Note that the Im $\bar\chi$, as a function of $\omega$, is proportional to the length of
the arcs for which
$(E_{\bf k}+E_{\bf Q-k})=\omega$. The lower and the upper
cutoffs are defined by the arcs touching the diamond, and the peak is contained
between these two cutoffs. As
$\omega$ increases, the arc length increases approximately linearly within the
diamond, and then drops approximately linearly. Moreover,
because the superconducting region around the Fermi line is very narrow, the
width of this peak is also very narrow, as in experiments. The peak is at $T_J({\bf k}_0)$, where ${\bf k}_0$ is located at the center of the diamond. This elementary
analysis  substantiates the numerical results in which none of the above
simplifying approximations were made.  

To understand the dependence of the scattering intensity on the momentum 
transfer perpendicular to the plane, it is necessary to consider the mixing of
the electronic wavefunctions between the layers. This cannot be a coherent 
superposition of states, otherwise the bands will be split, which, as stated
earlier, is not observed in experiments. However, a virtual mixing of the states
due to second order processes, similar to superexchange, is possible.
This virtual mixing can be described by approximately constructing the
operators\cite{Anderson2}: $\alpha_{{\bf k},\sigma,1}={(c_{{\bf k},\sigma,1}+
\eta({\bf k})c_{{\bf k},\sigma,2})/ \sqrt{1+\eta^2({\bf k})}}$, and
$\alpha_{{\bf k},\sigma,2}= {(c_{{\bf k},\sigma,2}+
\eta({\bf k})c_{{\bf k},\sigma,1})/ \sqrt{1+\eta^2({\bf k})}}$,
where $\eta({\bf k})$ is the mixing parameter. The order parameter that 
takes the mixing into account  is $\Lambda_{\bf k}=\langle \alpha^{\dagger}_{{\bf
k},\uparrow,1}\alpha^{\dagger}_{{-\bf k},\downarrow,1}+\alpha^{\dagger}_{{\bf
k},\uparrow,2}\alpha^{\dagger}_{{-\bf k},\downarrow,2}\rangle$.
It is also possible to rewrite this in terms of the operators $c^{\rm e}_{{\bf
k},\sigma}={1\over \sqrt{2}}(c_{{\bf k},\sigma,1}+c_{{\bf k},\sigma,2})$ and
$c^{\rm o}_{{\bf
k},\sigma}={1\over \sqrt{2}}(c_{{\bf k},\sigma,1}-c_{{\bf k},\sigma,2})$. Thus,
\begin{equation}
\Lambda_{\bf k}={(1+\eta({\bf k}))^2\over 1+\eta^2({\bf k})}\langle c^{{\rm
e}\dagger}c^{-{\rm e}\dagger}\rangle+ {(1-\eta({\bf k}))^2\over 1+\eta^2({\bf
k})}
\langle c^{{\rm o}\dagger}c^{-{\rm o}\dagger}\rangle,
\end{equation}
where we have used the notation $c^{{\rm e}\dagger}\equiv c^{{\rm
e}\dagger}_{{\bf k},\uparrow}$, $c^{-{\rm e}\dagger}\equiv c^{{\rm
e}\dagger}_{-{\bf k},\downarrow}$, etc.

This virtual mixing of the wave functions on the layers is a necessary
concomitant of the source of the pairing energy in the hopping terms in the
original Hamiltonian $-\sum_{{\bf k},\sigma,i\ne j}t_{\perp}({\bf k})(c_{{\bf
k},\sigma,i}^{\dagger}c_{{\bf k},\sigma,j}+{\rm h. c.})=-\sum_{{\bf
k},\sigma}t_{\perp}({\bf k})(n_{{\bf k},\sigma}^{\rm e}-n_{{\bf
k},\sigma}^{\rm o})$, where $n$ are the number operators. The pairing energy must be numerically augmented by its
expectation value. However, the two-particle tunneling term alone does not give
quite enough difference in occupancy to explain the experimental modulation
with $q_z$. In a forthcoming paper\cite{Anderson2}, we will show that this
tendency is strongly enhanced by including the ``superexchange" type terms
caused by virtual particle-hole hopping. In fact, it may be that the pairing
occurs primarily in the ``even" linear combination alone\cite{Symm}.

Then, any electron excited from
the condensate must leave behind a hole in the even state. Similarly, it can
accept only an electron in the odd state, hence the striking ${\rm
even}\leftrightarrow{\rm odd}$ selection rule observed in the neutron
scattering experiments. Note that although the odd states are nominally
unpaired, the addition or removal of an electron from the odd state destroys
the pairing criterion and costs one unit of pairing energy. Thus, there is a
gap for all one-electron excitations, even though only the even state is
paired.

In conclusion, we find that these neutron results give us a surprisingly direct
and complete picture of the nature of the pairing responsible for
superconductivity, not only showing us the symmetry and approximate form of the
energy gap, but even fixing the nature of the gap equation and the source of
the pairing energy. In particular, the strong correlation between the layers
excludes any purely intralayer mechanism for superconductivity, under our
assumption --- the singlet, and very plausible on other grounds --- that the
quasiparticle pairs do not have strong final state interactions.

This work was supported by the National Science Foundation: Grant Nos. 
DMR-9531575 and DMR-9104873. We thank B. Keimer for many stimulating
discussions.

\end{document}